**Title:** Seasonal and Secular Periodicities Identified in the Dynamics of US FDA Medical Devices (1976-2020): Portends Intrinsic Industrial Transformation and Independence of Certain Crises

**Author:** Iraj Daizadeh, PhD, Takeda Pharmaceuticals, 40 Landsdowne St. Cambridge, MA, 02139, iraj.daizadeh@takeda.com

**Abstract:**

The US Food and Drug Administration (FDA) regulates medical devices (MD), which are predicated on a concoction of economic and policy forces (e.g., supply/demand, crises, patents), under primarily two administrative circuits: Premarketing Notifications (PMN) and Approvals (PMAs). This work considers the dynamics of FDA PMNs and PMAs applications as an proxy metric for the evolution of the MD industry, and specifically seeks to test the existence (and, if so, identify the length scale(s)) of economic/business cycles. Beyond summary statistics, the monthly (May, 1976 to December, 2020) number of observed FDA MD Applications are investigated via an assortment of time series techniques (including: Discrete Wavelet Transform, Running Moving Average Filter, Complete Ensemble Empirical Mode with Adaptive Noise decomposition, and Seasonal Trend Loess decomposition) to exhaustively seek and find such periodicities. This work finds that from 1976 to 2020, the dynamics of MD applications are (1) non-normal, non-stationary (fractional order of integration < 1), non-linear, and strongly persistent (Hurst > 0.5); (2) regular (non-variance), with latent periodicities following seasonal, 1-year (short-term), 5-6 year (Juglar; mid-term), and a single 24-year (Kuznets; medium-term) period (when considering the total number of MD applications); (3) evolving independently of any specific exogeneous factor (such as the COVID-19 crisis); (4) comprised of two inversely opposing processes (PMNs and PMAs) suggesting an intrinsic structural industrial transformation occurring within the MD industry; and, (6) predicted to continue its decline (as a totality) into the mid-2020s until recovery. Ramifications of these findings are discussed.

**Introduction**

The history of the United States (US) medical device (MD) industry is one of innovation – a complicated evolution encompassing a very broad everyday (e.g., general purpose thermometers) and specialized (e.g., human-embeddable systems) medical products. At some point in time, each registered MD – no matter how menial from today's vantage point – is an outcome of certain investments. On a company or sector level, these investments have not only included the demand / supply side variables (such as the various people, processes, and systems required of a research and development firm to idealize, actualize, market, and secure economic rents from the sale of a MD product) but also to meet the national policies enforced by one or more national health bodies to ensure the MD's safe intended use [1]. Thus, an important assignment would be to identify and investigate metrics of output that may be used as proxies to track certain aspects of the sector, including its innovativeness and its general health.

Globally, the MD development process requires supervision and registration with a local health agency, which in the US would be the Food and Drug Administration (FDA)'s Center for Devices and Radiological Health (CDRH)[1]. Starting in 1976, CDRH received congressional mandates to ensure the safe and appropriate use of MDs via the Medical Device Amendments to the Federal Food, Drug, and Cosmetic (FD&C) Act[2], and subsequent legislation (see Table 1).

---

[1] Here, CDRH, FDA, or Agency may be used interchangeably.
[2] Pub. L. 94-295 enacted on May 28, 1976

**Table 1: Some Milestones in FDA Device Regulation since 1970)[3]**

| Year | US Drug Regulation |
|------|--------------------|
| 1970 | Cooper Committee is established, which "recommended that any new legislation be specifically targeted to devices because devices present different issues than drugs" |
| 1976 | Medical Device Amendments to the Federal Food, Drug, and Cosmetic (FD&C) Act |
| 1990 | Safe Medical Devices Act (SMDA) |
| 1992 | Mammography Quality Standards Act (MQSA) |
| 1997 | FDA Modernization Act (FDAMA) |
| 2002 | Medical Device User Fee and Modernization Act (MDUFMA) |
| 2007 | FDA Amendments Act (FDAAA), MDUFA II |
| 2012 | FDA Safety and Innovation Act (FDASIA), MDUFA III |
| 2016 | 21st Century Cures Act |
| 2017 | FDA Reauthorization Act (FDARA), MDUFA IV |

In total, the above laws allowed the FDA to develop regulations offered an opportunity to regulate the industry while simultaneously promoting its development through applying a classification scheme to MDs based on patient risk via the intended use of a given MD. Simplistically, the greater the risk (and thus higher the class), the most stringent the requirements for receiving registration to market a specific MD in the US[4]. In a great part, this risk-based approach led to two key regulatory registration paths: the Premarketing Notification (PMN; otherwise, known as the 510(k)) and Premarketing Approval (PMA). The PMN process is relatively administratively simpler and results in a clearance [2], while that of the PMA is (typically) more complex (as it may require clinical data) and results in formal approval from the Agency[5,6]. One must keep in mind the importance that both circuits include an application from the Sponsor seeking registration: either a PMN or PMA. Thus, the application represents the sponsor's assertion of the merits of the MD and its potential viability in the marketplace. In many ways, the act of submitting the application for registration is the sponsor's belief that all of the various inputs

---

(investments) have cumulated into an innovative product of interest to the marketplace. This is particularly true if the product attained registration status. Cumulatively, therefore, the total number of PMN and PMA Applications (and their sum Total MD (PMA+PMN)) over time would be a key piece of evidence (metric) supporting the evolution of innovativeness and/or other economies associated with the MD industry.

In this work, the 3 metrics of the regulated MD industry are considered: the number of PMN, PMA, and Total MD Applications. It is hypothesized that these metrics would be behave similarly to those of other economic variables, as each MD (and thus as a collective) is a resultant composite of various input – including those of the firm (e.g., people, processes, systems), those of the sector (e.g., supply/demand mechanics), as well as those of national policy and enforcement through a regulatory body. Unlike other economic variables, however, to the author's knowledge, little is known about the dynamics of these metrics and their potential importance to understanding the various factors that may have influenced their evolution (including its forecast). Importantly, in this case of MD development and the metrics selected, these factors include substantive economic activities (e.g., as crises) and/or health policy (e.g., laws) considerations.

Specifically, the focus of this work is on one key characteristic of economic variables (particular those for which sufficient longitudinal data is available) is the appearance of so-called economic or business cycles in the FDA regulated MD industry. Cycles are generally described in terms of wave mechanics in which a noticeable peak eventually leads to an trough and recycles – where the peak would be considered the pinnacle of economic prosperity (e.g., expansion) of some sort whereas the trough would be a temporally associated misery in productivity (e.g., contraction). From a certain perspective of exogenic strength, the ebb and flow of the variable would correspond to the time-varying strength of forces pressing on the metric. At least four broadly canonical cycles exist (beyond that of seasonal effects), although there has been advancements (see, e.g., [3]) summarized as:

- Kitchin Short-Term Cycle [4]: 3.5 years in length. Derived as a generalization "supported by a wide range of, annual statistics for Great Britain and the United States, and especially by monthly statistics of clearings, commodity prices, and interest rates for the two countries (page 10)." Kitchin writes that he agrees with a "Mr. Philip Green Wright when he suggests: 'Business and price cycles are due to cyclical recurrences in mass psychology reacting through capitalistic production. The rough periodicity of business cycles suggests the elastic recurrence of human functioning rather than the mathematical precision of cosmic phenomena (page 14).'"

- Juglar Mid-Term Cycle [5, 6]: 6-7 years in length with a 1-2 year precipitous drop. Besomi ([5], page 3) captures Juglar's thoughts that – based on banking, population, price of corn, import and exports, rents and public revenue statistics across England, US, Prussia and Hamburg – there was a "a strict correlation … and that changes go through specific phases, always the same, and are in concordance in the countries where commerce and industry are more development. From this regularity, Juglar inferred that the common premise to all crises lies in the excesses of speculation and in the inconsiderate expansion of industry and trade (*ibid*, page 4)."

- Kuznets Medium-Term Cycle [7]: 15-25 years ([8] stated 15-20 years; Kuznets specified approximately (but equal to or greater than) 20 years (see Tables 3 and 4 on pages 204 and 205, respectively in Kuznets, 1930 across US and Europe and various goods and services (including with caveat trusts)). Abramovitz [8] nicely summarizes this perspective in trichotomized phases: a rebound from depression ("growth rate of output was accelerating to maximum (page 351)"), steady growth ("smoothed growth rate was high enough to keep the labour force well employed. It was interrupted by short mild recessions, but at cyclical peaks the demand for labour pressed on supply (351/352))", followed by a depression or stagnation ("actual output always fell sharply; smoothed output usually declined or at best grew very slowly (page 352))."

- Kondratieff Long-Term Cycle ([9]:  50 years (+/- 5-7 years (*ibid*, page 111)). Kondratieff derives 3 cycles each roughly 50 years (more or less) across a series of econometrics across France, England, Germany, the US, and the "whole world" (*ibid*, Table 1, page 110). Importantly, the author concludes the following proposals: (1) "long waves below … to the same complex dynamic process in which the intermediate cycles of the capitalistic economy with their principal phases of upswing and depression run their course (*ibid.* page 111);" (2) "during the recession of the long waves, agriculture, as a rule, suffers an especially pronounced and long depression (*ibid*);" (3) "during the recession of the long waves, an especially large number of important discoveries and inventions in the technique of production and communication are made, which, however, are usually applied on a large scale only at the beginning of the next long upswing (*ibid*);" (4) "at the beginning of the upswing, gold production increases as a rule… (*ibid*);" (5) "It is during the period of the rise of the long waves, i.e., during the period of high tension in the expansion of economic forces, that, as a rule, the most disastrous and extensive wars and revolutions occur (*ibid*)."

Here, the key hypothesis that is tested is: assuming PMN, PMA and the Total MD Applications are a proxy metric associated with the MD industry (and assuming therefore these variables act as other econometrics), do latent periodicities exists? If so, what are the time lengths of such periodicities. The hypothesis is tested via several statistical approaches, based on two objectives: (1) to understand the intrinsic nature of the 3 time series (viz., descriptive statistics) and, based on this information, (2) to resolve any identified periodicities accordingly. The statistical routines used to describe:

- The data include typical distribution statistics (e.g., $1^{st}$, $2^{nd}$ and higher moments), normality, seasonality, linearity, stationarity, long-range dependency, and structural break.

- The periodicities include Refined Moving Average Filter (RMAF), Seasonal Trend Loess (STL), wavelet power spectra, and the Complete Ensemble Empirical Mode with Adaptive Noise decomposition (CEEMDAN).

An explanation of each of the algorithms and why they were selected are part of the Materials and Methodologies section. Thereafter, *prima facie* results are presented. The manuscript closes with an interpretation of the results and key conclusions including limitations of the study and future directions for continued research.

**Materials and Methodologies**

While details of the materials (including data acquisition and preparation) and methodologies (including R programming code) are presented in the accompanying Supplementary Materials as a means to fully replicate and/or extend this analyses, this section summarizes the data sources and its preparation, as well as the rationale and statistical methodologies used in performing the analyses.

*Data Sources and Data Preparation*

The data was focused on applications (and not registrations) as the key hypotheses surrounding efficiencies associated with the MD industry (and not, e.g., those of the FDA registration process). The US FDA data is considered in this report as the 'authorized system of record;' thus, PMN and PMA data was obtained from the US FDA repository, as there is no known repository containing failed (that is, non-authorized for sale) MDs.

- PMNs: The data were obtained from the FDA site: https://www.fda.gov/medical-devices/510k-clearances/downloadable-510k-files on June 30, 2021. The files included PMN7680.ZIP (1976-1980), PMN8185.ZIP (1981-1985), PMN8690.ZIP (1986-1990), PMN9195.ZIP (1991-1995), and PMN96CUR.ZIP (1996-Current).

  - Date Range: May, 1976 to Dec, 2020

  - Total Number of Records: 158,961

- PMAs: The data were obtained from the FDA site: https://www.fda.gov/medical-devices/device-approvals-denials-and-clearances/pma-approvals#pma (under section "PMA/PDP Files for Downloading" on June 30, 2021. The files included pma.zip, "which contains information about the releasable PMAs (ibid)."

- Date Range: Oct., 1960 to Dec., 2020 (Note: The data was truncated to May 1976 to Dec 31, 2020 to allow direct comparison of the earliest PMN record. A negligible deletion of 178 records.)

- Total Number of Records: 44,831 (44,805 with the truncation)

These data sources were culled for "DATERECEIVED" (Application); that is, the date the application was received by FDA; and imported into Excel, wherein the dates were counted on a monthly scale and then exported as Comma-Separated Values (CSV) file for input into the R programming environment.

In total, 3 variables comprised the complete dataset: PMN Applications, PMA Applications, and Total MD Applications (that is, the monthly number of PMNs and PMAs were simply summed) – each with 536 values (the sum of all observations within a given month from May 1976 to December, 2020). To summarize, the 3 time series were:

- Time Series #1: PMN Applications: MDs seeking PMN (510(k)) registration.

- Time Series #2: PMA Applications: MDs seeking PMA registration.

- Time Series #3: Total MD Applications: MDs seeking either PMN or PMA registration.

*Statistical Analyses*

The general intent of the statistical analyses were two-fold: (1) to understand the intrinsic nature of the 3 time series (viz., descriptive statistics) and, based on this information, (2) to resolve periodicities accordingly. Note: As discussed further below, certain data attributes elucidated from certain tests necessitated further analysis (see Results) specifically around non-stationarity and long-term memory.

There are many statistical approaches with a capability to characterize a given dataset including decomposition (viz., reduction to seasonal, trend, and random (stochastic) contributions and inversely reconstructing the time series (within some sort of acceptable error) through some additive or multiplicative combination), structural changes (viz., identification of meaningful changes in certain distribution attributes), data (e.g., correction denoising and/or missing data), and dimensionality reduction (e.g., techniques to reduce or identify the variables that would represent key properties of the original variable space) and so on. Here, the algorithms selected were a result of: appropriateness based on the time series structure (e.g., non-linearity and non-stationarity), accessibility to the algorithm (access via the R Project), as well as the nature of the signal to be resolved (periodicity). Thus, an effort has been made to use known methodologies (where possible) and cross-validating the results through either using different approaches (ideally with limited theoretical overlap) or exploring the parameter space of a given algorithm. As this work is a result of applying known methodologies, all supportive mathematical formulae are deferred via citation. Unless specified otherwise, all methods presented followed standard implementation and default parameters were used (as appropriate) throughout the analyses.

Step 1: Statistical characteristics of the data

This step simply explores the distribution of the data from a time series perspective, estimating its general characteristics (e.g., moments) as well as outlining its dynamics (e.g., its stationarity and long-range dependency (LRD)). Either the characteristics of the distribution or properties of the dynamics may alter the calculations, since – for example –a stationary or non-LRD time series may allow for 'simpler' approaches to the analysis, as the moments would be time invariant or individual signals separable, respectively. The analyses followed the following prescription:

Time series loaded and descriptive statistics performed [10; 11: R Package: 'fBasics'; 12: R Package: 'forestmangr']: In this step, the data was read as a time series into the R program, and descriptive statistics were assessed via the following tests:

- Normality [13: R package: 'foreach'; 14: R package: 'nortest']: Anderson-Darling (A-D), Cramer-von Mises (CvM), and Lilliefors (Kolmogorov-Smirnov) (K-S) normality tests

- Seasonality [15: R package: 'seastests']: WO, QS, Friedman and Welch tests

- Nonlinearity [16: R package: 'nonlinearTseries']: Teraesvirta's and White Neural Network tests, and Keenan, McLeod-Li, Tsay, and Likelihood Ratio tests

- Stationarity [17: R package: 'aTSA']: Augmented Dickey–Fuller (ADF), Kwiatkowski-Phillips-Schmidt-Shin (KPSS), and Phillips–Perron (PP) Unit Root Tests

- LRD: Qu and Multivariate local Whittle Score type (MLWS) tests [18: R package: 'LongMemoryTS'], autocorrelation function (ACF) [10: R package 'stats'], and Hurst Exponent [19: R Project: 'pracma' (hurstexp); 20: R Project: 'tsfeatures' (hurst)]

- Order of integration [18: R package: 'LongMemoryTS']: Geweke-Porter-Hudak (GPH) estimator of fractional difference

Given that 2 tests (viz., MLWS and Qu test) suggested 'spurious' LRD, yet the Hurst Exponent and the existence of non-zero/non-unity (fractional) order of integration existed; thus, statistical estimation of structural breaks was performed using the standard dynamic programming model of Bai and Perron as implemented by [21-23: R Project: 'strucchange'; 24: R Project: [tseries]. In this approach, the definition of structural break is one in which there is a some sort of significant change in the parameters of a (linear) regression model. The existence of breaks would strongly affect the selection of statistical algorithms.

Step 2: Statistical determination of periodicities latent in the data

Shorter-term Periodicities: Seasonal trend decomposition via Loess method (STL) [10: R Package: 'stats'], Refined Moving Average Filter (RMAF) [25: R Package: 'rmaf'], and the wavelet power spectrum using a Morlet wavelet under a smoothing (Loess) construction [26: R Package: 'WaveletComp'] were used to investigate the short-term structure of the time series data. For the latter, the average period versus the average power for each method was then calculated to elucidate the main periodicities (*ibid*). The dominant frequencies identified were then re-confirmed via spectral analysis [27-28: R Package: 'forecast']. This approach allowed for cross-validation as these methods are orthogonal – that is, there limited-to-no methodological overlap between the methods chosen.

Longer-Term Periodicities: STL, RMAF, and the Complete Ensemble Empirical Mode with Adaptive Noise decomposition (CEEMDAN) [29-30: R Project: 'Rlibeemd'] were used to determine the longer trend. The challenge of resolving longer periodicities were multi-factorial and rested with the non-stationary, non-linear, and multiple structural break nature of the data over the duration of the data series. Thus, the CEEMDAN method, which utilizes an adaptive decomposition, has been considered the method of choice to tackle such programs given its flexibility with this type of data [31].

**Results:**

*Objective 1: Statistical Properties of US MD Applications*

The evolution of PMN and PMA Applications seem to follow inverse trajectories, while that of Total MD Applications resembles the sum of the two qualitatively (Figure 1). The trendline for PMNs (Figure 1a) suggests a significant decay since the peak in the early 1990s, while for PMAs there has been an acceleration since 2000 (Figure 1b). While PMA Applications (Figure 1b) show a somewhat relative decline in peak in 2020, it is relatively small. The evolution of Total MD Applications is notable due to the clear presence of a single period, with a decline prior to the year of COVID-19 (2020). Note: The scales of the trendlines (Figure 1 - right in green) are slightly different than that of the original observations to better resolve the yearly distributions.

**Figure 1: Time Evolution of PMN Applications (Top), PMA Applications (Middle), and Total MD Applications (Bottom): Observed Number of Applications (Red); Estimated Trend (Left) and Estimated Trend Only (Right) (Refined Moving Average with a Period of 12 Months)**

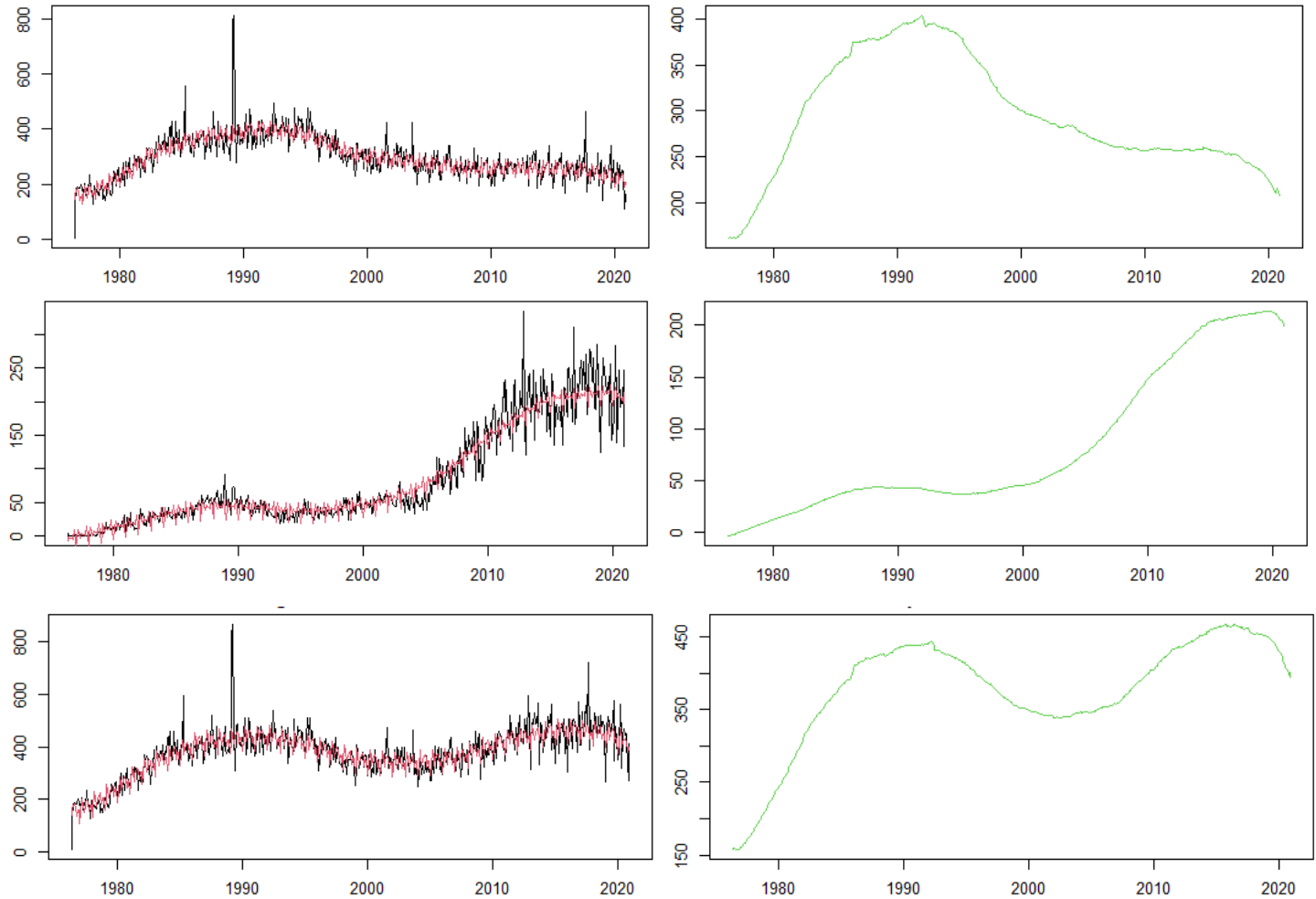

Shifting our attention to the distribution properties, Tables 2, 3 and Figures 2 presents the results of the various tests and finds that all three time series are non-normal (skewed with differences in tail thickness: PMN-leptokurtic , PMA-mesokurtic, and total MD- platykurtic relative to a normal distribution, but similar in spread), non-stationary, seasonal, non-linear, with considerable long-memory (see Figure 2 in which there is a long decay to zero) with fractional order of integration, significant persistency, and the existence of structural changes.

**Table 2: Summary Statistics of US FDA MD Applications (Total Number of Observations=536 per Time Series; Rounded to 2 Significant Digits; Units in Months)**

| Statistic | PMN Applications | PMA Applications | Total MD Applications |
|---|---|---|---|
| Minimum | 3 | 0 | 7 |
| Maximum | 813 | 335 | 869 |
| 1st Quartile | 247 | 32 | 327 |
| 3rd Quartile | 347.25 | 135.25 | 437.25 |
| Mean | 296.11 | 83.59 | 379.7 |
| Median | 286 | 50 | 388 |
| Standard Error (Mean) | 3.45 | 3.25 | 4.11 |
| Lower Confidence Limit (Mean) | 289.33 | 77.21 | 371.62 |
| Upper Confidence Limit (Mean) | 302.89 | 89.98 | 387.78 |
| Variance | 6389.99 | 5665.03 | 9058.6 |
| Standard Deviation | 79.94 | 75.27 | 95.18 |
| Skewness | 1 | 1.06 | -0.02 |
| Kurtosis | 5.14 | -0.11 | 2.34 |
| Total Records | 158714 | 44805 | 203519 |

**Figure 2: Auto (Serial) Correlation Function Versus Lag (Months): PMN Applications (Top), PMA**

**Applications (Middle), and Total MD Applications (Bottom) [95% Confidence Levels Denoted in Blue]**

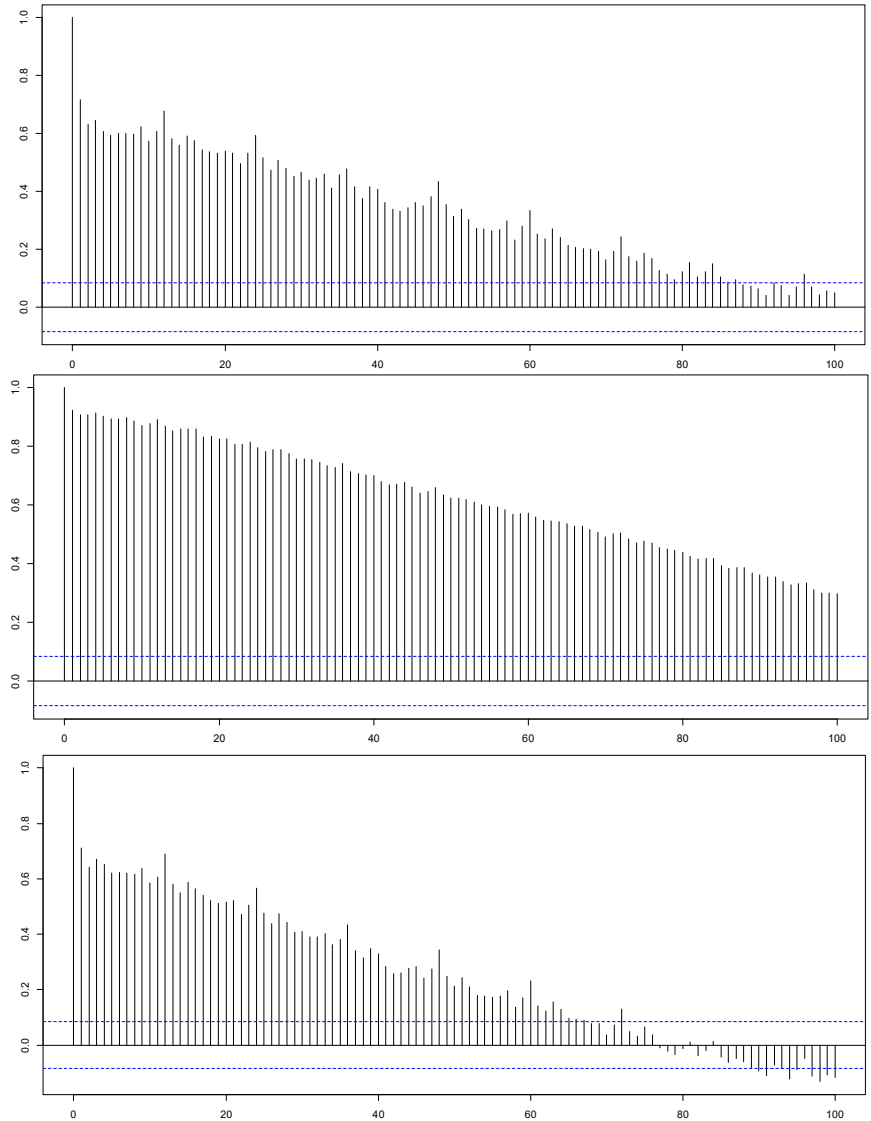

**Table 3: Summary of Normality, Stationarity, Seasonality, Long-memory, and Non-linearity Test Results of US FDA MD (Rounded to Tenths; Units in Months; Rejection of the Null Hypothesis was Based on p-Value < 0.01; Results are Presented in Supplementary Materials)\***

| Test Category | Tests* | Test Result Against Null | | |
|---|---|---|---|---|
| | | **PMN Applications** | **PMA Applications** | **Total MD Applications** |
| **Normality** | A-D, CvM, KS | Reject normality | Reject normality | Reject normality |
| **Seasonality** | WO, QS, Friedman | Seasonality | Seasonality | Seasonality |
| **Linearity** | TNN, WNN | Reject linearity+ | Reject linearity | Reject linearity |
| **Stationarity** | ADF, KPSS, PP | Reject stationarity | Reject stationarity | Reject stationarity |
| **Order of Integration (fractional differencing order *d*)** | GPH | 0.39 | 0.65 | 0.44 |
| **Long-memory** | ACF | Yes | Yes | Yes |
| **Hurst Exponent** | 1) Simple R/S Hurst Estimate 2) 0.5 plus the maximum likelihood estimate of the fractional differencing order d# | 1) 0.83 2) 0.93 | 1) 0.86 2) 1.0 | 1) 0.77 2) 0.92 |
| **Structural Breaks** | Significance testing of EFP with OLS-CUSUM, OLS-MUSUM, Rec-CUSUM, and Rec-MOSUM$ | Reject no structural changes | Reject no structural changes | Reject no structural changes |

\* A-D: Anderson-Darling; CvM: Cramer-von Mises; KS: Lilliefors (Kolmogorov-Smirnov); ADF: Augmented Dickey-Fuller; KPSS: Kwiatkowski-Phillips-Schmidt-Shin; PP: Phillips-Perron MLWS: WO: Webel-Ollech; TNN: Teraesvirta Neural Network; WNN: White Neural Network; GPH: Geweke and Porter-Hudak; ACF: autocorrelation function

+ Null hypothesis of linearity (in 'mean') rejected at the p-value < 0.085 (TNN) and 0.0087 (WNN)

\# Calculation is difference than that above, see Haslett and Raftery, 1989. Generally, the Hurst Exponent is related to the fractional dimension, *d*, by the equation: *d=2-Hurst*.

$ EFP: empirical fluctuation processes; OLS-CUSUM: ; OLS-MUSUM: ; Rec-CUSUM: ; Rec-MOSUM:

*Objective 2: Periodicity Latent within US MD Applications*

*Short-term Cyclicity:* STL, RMAF, the wavelet power spectrum using a Morlet wavelet, and spectral analysis reconfirmed seasonality as well as elucidated short-term periodicity. Seasonality (Figure 3) pictographs suggest multiple short-term cyclicity at the 1-year mark or less; spectral analysis resolved dominant peaks at 1 year (PMN), third-year (PMA), and quarter-year (TotalMD), seemingly mapping against business quarters. Where red represents increased foci of energy, the wavelet power spectra (Figure 4) presents conceptually near similar results, with a 1 year period or less oscillating over the full reporting period for all three time series. Of interest, there is a concentration of energy (red) around 1 year from 1985-1990 for PMNs, 2003-2020 for PMAs, and similarly both 1985-1990 and 2003-2020 for Total MDs.

*Longer-Term Cyclicity:* RMAF (Figure 1), and CEEMDAN (Figure 5) algorithms elucidated longer trends. Its challenging to view a periodic structure in the RMAF with the exception of Total MD, in which a clear single periodic structure (with two peaks and a trough) is resolved (Figure 1 bottom). The two respective peaks are located at: April, 1992, July 2016, respectively; a period of 24 years and 3 months. The result of the CEEMDAN methodology depicts peaks at around 5 years across time-lengths and time series data.

**Figure 3: Seasonal Periodicity (via STL) for PMN (Top), PMA (Mid), and Total MD Applications (Bottom)**

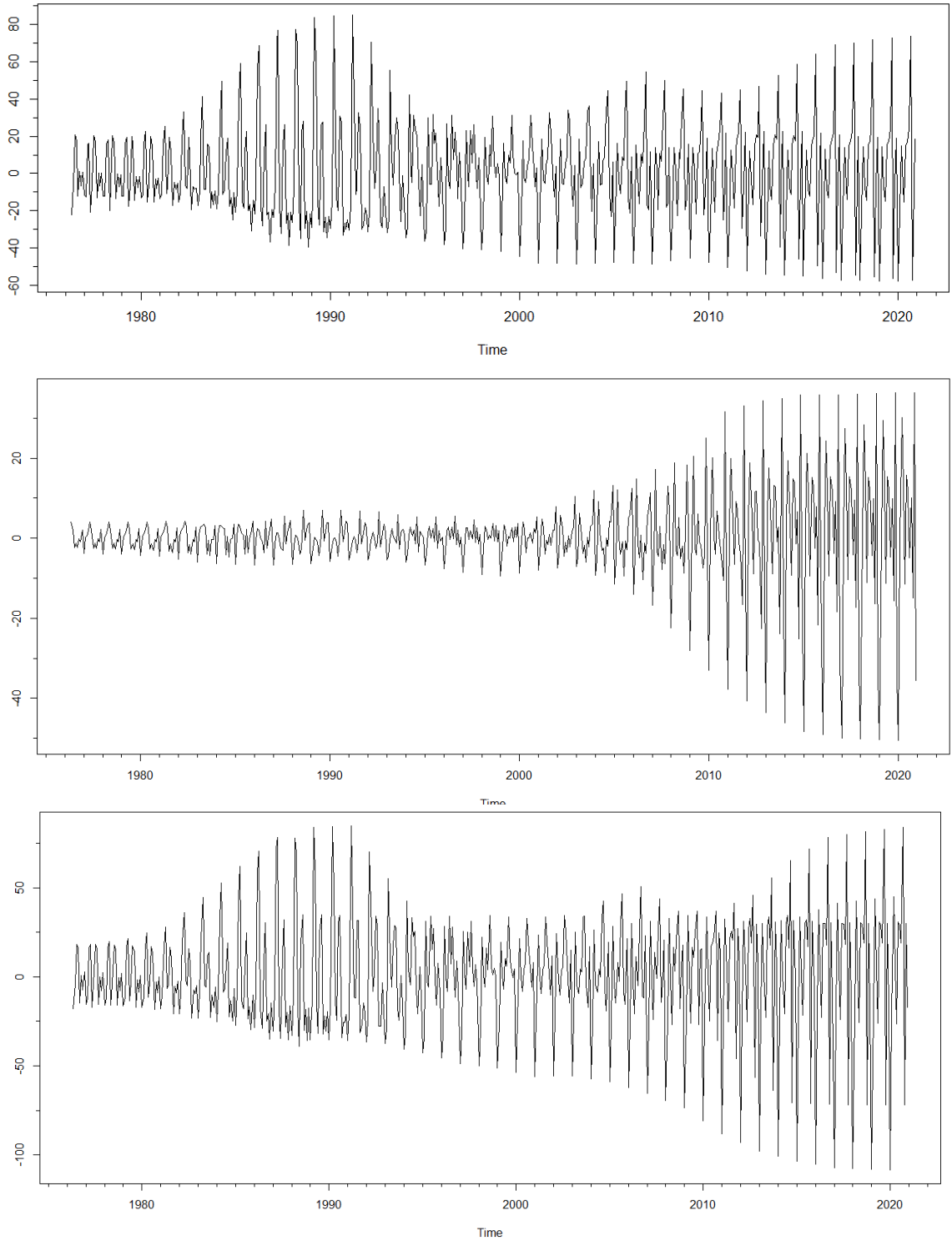

**Figure 4: Wavelet Power Spectra for PMN (Top), PMA (Middle), and Total MD (Bottom) Applications**

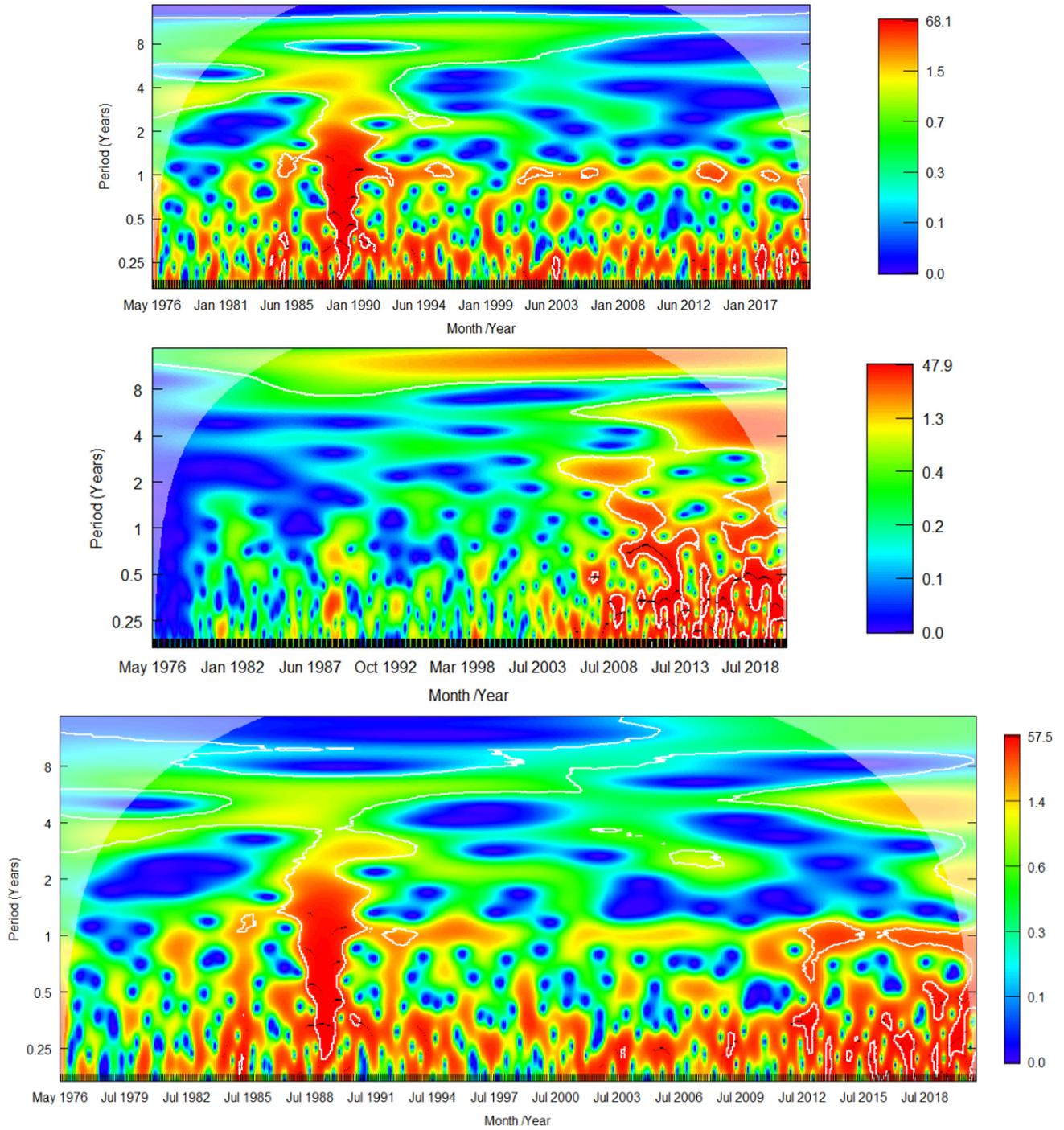

**Figure 5: CEEMDAN trends for PMN (Top), PMA (Mid), and Total MD (Bottom) Applications**

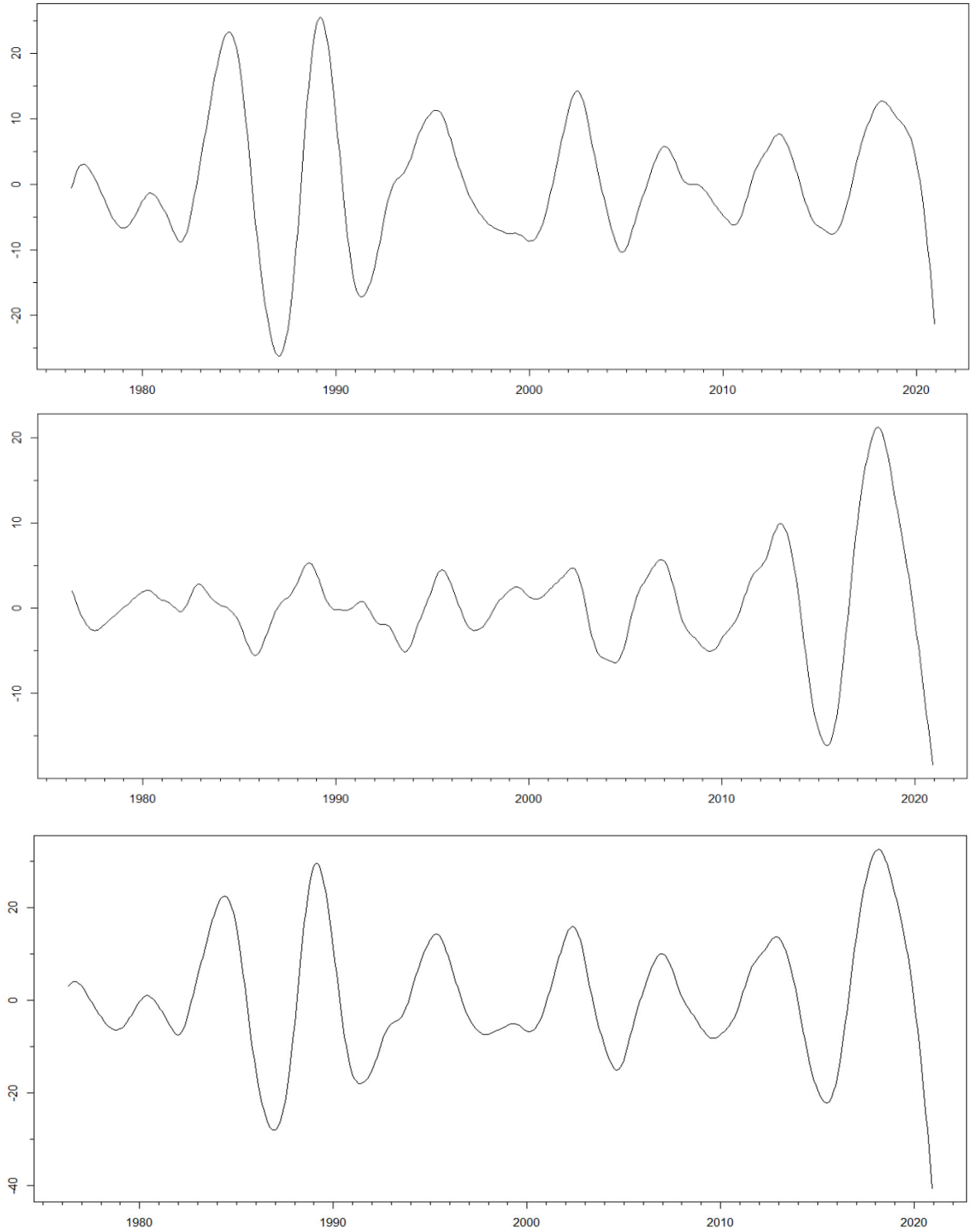

A summary of the periodicities for ease of reference along with the sources is presented in Table 4:

**Table 4: Mapping of Broad Canonical Economic Cycles with that of Periodicities Associated with FDA Medical Devices (Units in Years) as Identified in this Study (see Text for Details)**

| Theory | Canonical Periodicity | PMN | PMA | Total MD |
|---|---|---|---|---|
| Seasonal/Yearly Cycle* | 0.25/1 | 1 | 0.3 / 1 | 0.25 / 1 |
| Kitchin Short-Term Cycle | 3.5 | | | |
| Juglar Mid-Term Cycle# | 7-11 | 5-6 | 5-6 / 8 | 5-6 |
| Kuznets Medium-Term Cycle^ | 15-25 | | | 24 |
| Kondratieff Long-Term Cycle | 40-60 | | | |
| Reference: | | | | |
| * Table 3, Figure 3 and 4, and dominant peaks of spectral analysis (see Supplementary Materials) # Figure 4 (middle), Figure 5 ^ Figure 1 (Total MD) (as well as CEEMAD (see Supplementary Materials)) | | | | |

**Discussion and Conclusion**

This work concerns itself with the FDA regulated MD industry and select metrics (PMN, PMA, Total MD) that may be used to explore its evolution. The behavior of the proposed metrics are presumed to be similar to that of other econometrics (e.g., labor, pricing, and production), given the diversity of inputs of varying strengths used to develop a specific MD (output). A specific property of econometrics is the presence of periodicities. This work continues to add support for the existence of such periodicities, as several were found via these proposed metrics. The robust finding of periodicities across a broad assortment of econometrics data (including that of FDA-regulated medicines [32]) strongly suggests the existence of potential laws (akin to those identified in physical systems) that may reflect (or indeed govern) aspects of growth and ebb dynamics observed in these curvilinear structures. Future work may consider using these data (and/or those of FDA-approved medicines) to build such a theory, as the data are robust, easy to collect, and relatively granular (daily values that can be aggregated).

Additionally, this work also sought to identify the time lengths of the latent time series periodicities. Importantly, both seasonal and secular cyclicities were identified. These included: seasonal and yearly, mid-term (Juglar), and medium-term (Kuznets) cycles. A longer (Kondratieff) term (> 20) year cycle was not observed from the methods used. The seasonal/yearly periodicities as well as a Kuznets 24 year medium term cycles were the most easily elucidated, based on the selected methods; indeed, the Kuznet cycle was derived from simple observation (albeit much more clearly post-RMAF).

Theoretically, how would one translate the theorists conjectures of periodicities to the MD industry? For the medium-term (24-year) cycle, and leveraging Kuznets theory, the author speculates that the existence (and use) of the substructures of the PMN and PMA Application curves (metrics) give us a unique insight into the MD industry from a periodicity perspective. Apparently from the simple PMN and PMA plots (Figure 1), it would seem that the industry is undergoing a potential transformation. The number of PMNs since at least 2000 has been stagnate to trending downward from a relative peak in the early 1990s, while PMAs since 2000 has been clearly growing in a striking-cobra-like pattern. Taken together, the collective metric (Total MD) resembles a clear Bactrian-camel-like structure with two clear peaks and a trough in the-1990s and mid-2010s. This would suggest an industry oriented movement from simple (lower risk, lower class) MDs to ones that are more complex (higher risk, higher class). Entrepreneurial tendencies grew as of 2000 to build complex MDs (e.g., human embedded systems) requiring additional health authority review and oversight (PMA); presumably driving economic rents given the increase in production costs. PMN activity stagnated due to lack of innovative creativity. Unlike other industries, however, PMAs would not generally replace PMNs – that is, we still need thermometers; thus, there is a floor to PMN Applications, whereas there is no limit to those of PMAs.

The seasonal/1 year and Juglar cycles are also of considerable interest. The latter specifically as the 5-6 year cycle was persistent via both PMAs and PMNs and throughout the data reporting duration of 44 years. An explanation for these is outstanding but may be speculated to reflect time lengths required for implementing simple to moderate innovation design changes. Imagine a MD in which a specific correction or addition to functionality was made. The updated (new) MD would be subsequently tested, placed into production, an application sought and registration granted by the HA. The rate of MD development in this context would be relatively much shorter than an industry transformation.

As noted by the periodicity theorists, there is little impact of crises to long-term tendencies. Consistent with the medium-term findings of Kuznets, there is no obvious impact of economic crises on Total MD Applications. There was a subsequent decrease in Total MDs prior to the recent severe acute respiratory syndrome coronavirus 2 (SARS-CoV-19; HCoV-19; COVID-19) crisis. In fact, there has not been any noticeable change at least in this study due to the crises on the cycles; Daizadeh, 2020, also did not notice any blatant impact of the COVID-19 crisis on medicines development at least as of Aug 2020 (Daizadeh, 2021b). it would be anticipated from the data that Total MD Application will continue a downward ascent until the mid-2020s prior to rising again, with a potential drop of at least 25% anticipated. PMNs will continue to drift seemingly. It therefore seems reasonable that PMAs would fall, assuming the continued structure.

Lastly, this work also has provided insight into the data themselves. We learn that the data is non-normal, non-linear, and non-stationary with specific characteristics (lopsided and fat tailed). We also learned that there is an intrinsic long-range dependency (LRD) reflecting memory dynamics as well as multiple structural breaks. Both of these features suggest statistical avenues to generate and investigate hypotheses related to exploring the impact of specific exogenous influences. The Chronological Hurst Exponent, which algorithmically leverages LRD, and Structural Breakpoint Analysis has been used in this way for FDA US medicines as an attempt to link economic events (e.g., crises) and policy interventions to changes in either LRD or structural breaks accordingly [34] – a similar experiment could be performed for MDs [35, 36].

This work concludes that (1) PMA and PMN data may be viewed as a proxy measure of innovativeness and certain economies in the MD industry; (2) similar to other econometrics in that periodicities exist are present in these metrics; (3) seasonal/1-year, Juglar and Kuznets periodicities are present in the metrics; (4) these metrics do not seem affected by specific crises (such as COVID-19); (5) PMNs and PMAs evolve inversely and suggest a structural industrial transformation; (6) Total MDs are predicted to continue their decline into the mid-2020s prior to recovery (thus, these metrics may play a greater role in predicting the evolution of the MD industry).

## Acknowledgements

The author extends gratitude N.D., S.L.D., and N.L.D. for their support of the manuscript.

## Disclosures

The author is an employee of Takeda Pharmaceuticals; however, this work was completed independently of his employment. See Supplementary Materials for all data and methods to replicate (or extend) the results presented herein.

# SUPPLEMENTARY MATERIALS

citation()

 R Core Team (2020). R: A language and environment for

 statistical computing. R Foundation for Statistical

 Computing, Vienna, Austria. URL

 https://www.R-project.org/.

version

            _

platform     x86_64-w64-mingw32

arch       x86_64

os        mingw32

system      x86_64, mingw32

status

major      4

minor      0.3

year      2020

month      10

day      10

svn rev        79318

language       R

version.string R version 4.0.3 (2020-10-10)

nickname       Bunny-Wunnies Freak Out

#Read in file and process

MD<-read.csv(XXX)

str(MD)

MD

   PMN PMA TotalMD

1    3   4      7

2  128   0    128

3  184   0    184

4  189   0    189

5  183   0    183

6  184   0    184

7  200   0    200

8  184   3    187

9  190  1   191

10  164  0   164

11  205  0   205

12  159  0   159

13  186  3   189

14  184  0   184

15  169  1   170

16  233  0   233

17  175  0   175

18  176  1   177

19  128  0   128

20  196  0   196

21  145  0   145

22  160  0   160

23  179  0   179

24  169  4   173

25  160  2   162

26  193  2   195

27  193  1   194

28 171 3   174

29 147 0   147

30 158 3   161

31 142 6   148

32 169 8   177

33 203 4   207

34 180 7   187

35 189 5   194

36 168 4   172

37 160 7   167

38 173 20   193

39 245 10   255

40 241 13   254

41 195 13   208

42 222 16   238

43 244 11   255

44 262 6   268

45 189 13   202

46 238 9   247

47 216  7    223

48 280  18   298

49 240  19   259

50 194  19   213

51 282  9    291

52 221  13   234

53 263  23   286

54 295  21   316

55 268  22   290

56 238  31   269

57 245  6    251

58 243  27   270

59 305  16   321

60 284  18   302

61 285  36   321

62 281  14   295

63 301  17   318

64 292  21   313

65 227  9    236

66 290 12 302

67 262 11 273

68 247 14 261

69 247 22 269

70 242 23 265

71 329 40 369

72 321 28 349

73 330 33 363

74 328 29 357

75 295 17 312

76 309 19 328

77 243 16 259

78 284 29 313

79 301 20 321

80 352 36 388

81 307 12 319

82 276 23 299

83 375 29 404

84 320 36 356

85 325 34    359

86 355 29    384

87 337 29    366

88 382 21    403

89 384 25    409

90 327 23    350

91 348 36    384

92 338 43    381

93 400 29    429

94 408 30    438

95 357 31    388

96 427 34    461

97 317 33    350

98 307 43    350

99 429 30    459

100 356 27    383

101 341 26    367

102 341 27    368

103 395 31    426

104 300 32    332

105 345 27    372

106 367 40    407

107 374 31    405

108 554 40    594

109 369 31    400

110 362 30    392

111 362 26    388

112 370 41    411

113 318 28    346

114 314 44    358

115 319 36    355

116 382 31    413

117 301 28    329

118 355 29    384

119 355 50    405

120 372 37    409

121 333 24    357

122 355 44    399

123 362  31    393

124 373  47    420

125 382  44    426

126 391  49    440

127 290  46    336

128 354  47    401

129 318  35    353

130 349  60    409

131 407  54    461

132 341  48    389

133 318  52    370

134 365  57    422

135 342  55    397

136 462  60    522

137 365  49    414

138 374  54    428

139 301  57    358

140 430  47    477

141 341  55    396

142 337 51    388

143 412 56    468

144 386 60    446

145 341 68    409

146 315 50    365

147 388 42    430

148 415 63    478

149 345 46    391

150 333 42    375

151 329 61    390

152 287 91    378

153 403 47    450

154 423 46    469

155 784 42    826

156 813 56    869

157 424 46    470

158 277 32    309

159 415 69    484

160 415 73    488

161 343 72   415

162 370 56   426

163 361 53   414

164 347 55   402

165 386 43   429

166 394 49   443

167 454 59   513

168 329 41   370

169 387 53   440

170 355 38   393

171 473 38   511

172 434 54   488

173 327 51   378

174 357 60   417

175 382 57   439

176 372 44   416

177 331 35   366

178 349 29   378

179 429 33   462

180 428  47    475

181 365  49    414

182 345  36    381

183 434  43    477

184 378  46    424

185 335  35    370

186 391  53    444

187 373  50    423

188 363  41    404

189 359  46    405

190 384  52    436

191 407  42    449

192 350  33    383

193 422  30    452

194 491  48    539

195 496  39    535

196 424  35    459

197 436  34    470

198 364  29    393

199 391  28    419

200 405  32    437

201 391  23    414

202 383  22    405

203 444  35    479

204 409  34    443

205 375  18    393

206 442  24    466

207 390  39    429

208 437  34    471

209 318  21    339

210 353  28    381

211 396  38    434

212 400  26    426

213 355  22    377

214 361  25    386

215 475  31    506

216 369  30    399

217 376  27    403

218 444  33    477

219 437  20    457

220 429  36    465

221 435  22    457

222 373  20    393

223 419  37    456

224 368  32    400

225 329  39    368

226 396  31    427

227 475  34    509

228 406  55    461

229 393  54    447

230 463  50    513

231 394  31    425

232 410  51    461

233 363  33    396

234 347  41    388

235 399  33    432

236 355  37    392

237 382  20    402

238 283  26    309

239 314  33    347

240 353  41    394

241 374  44    418

242 348  30    378

243 363  26    389

244 376  42    418

245 397  33    430

246 348  43    391

247 381  32    413

248 357  42    399

249 295  27    322

250 301  33    334

251 368  38    406

252 320  33    353

253 357  33    390

254 396  40    436

255 307  36    343

256 335  28    363

257 366  34    400

258 307  36    343

259 293  37    330

260 303  34    337

261 276  29    305

262 291  33    324

263 302  41    343

264 309  41    350

265 265  32    297

266 295  57    352

267 304  49    353

268 292  57    349

269 325  47    372

270 329  53    382

271 331  24    355

272 273  52    325

273 229  25    254

274 267  56    323

275 349  58    407

276 348  38    386

277 281  38    319

278 308  33    341

279 299  63    362

280 327  52    379

281 256  66    322

282 292  63    355

283 325  43    368

284 325  38    363

285 249  40    289

286 327  42    369

287 278  51    329

288 260  46    306

289 240  47    287

290 268  48    316

291 257  42    299

292 336  54    390

293 263  34    297

294 275  36   311

295 252  59   311

296 296  53   349

297 242  50   292

298 246  53   299

299 312  53   365

300 290  65   355

301 314  60   374

302 299  50   349

303 338  56   394

304 422  49   471

305 260  59   319

306 303  58   361

307 286  68   354

308 283  56   339

309 265  44   309

310 265  53   318

311 326  62   388

312 274  56   330

313 349 50　399

314 266 46　312

315 341 47　388

316 300 44　344

317 325 44　369

318 336 64　400

319 258 69　327

320 317 80　397

321 261 43　304

322 279 48　327

323 326 52　378

324 289 56　345

325 270 36　306

326 308 65　373

327 306 47　353

328 282 59　341

329 424 40　464

330 258 39　297

331 245 56　301

332 283  61   344

333 197  52   249

334 267  33   300

335 286  66   352

336 277  73   350

337 247  49   296

338 276  51   327

339 253  45   298

340 277  45   322

341 312  53   365

342 247  57   304

343 274  39   313

344 269  70   339

345 206  63   269

346 235  38   273

347 285  63   348

348 243  57   300

349 268  51   319

350 324  77   401

351 257 62    319

352 291 89    380

353 322 77    399

354 246 77    323

355 214 92    306

356 276 118   394

357 218 76    294

358 261 115   376

359 287 101   388

360 268 76    344

361 263 104   367

362 309 70    379

363 284 75    359

364 340 81    421

365 338 97    435

366 247 111   358

367 257 86    343

368 249 103   352

369 266 62    328

370 232  75    307

371 273 104    377

372 252  98    350

373 244 107    351

374 212  73    285

375 264  72    336

376 299 111    410

377 260 108    368

378 244 132    376

379 247  97    344

380 287 101    388

381 205  88    293

382 239  88    327

383 260 162    422

384 282 123    405

385 226 113    339

386 260 150    410

387 255 139    394

388 287 133    420

389 331 129    460

390 244 125    369

391 252 170    422

392 303 145    448

393 192 82    274

394 247 170    417

395 278 159    437

396 294 95    389

397 226 95    321

398 277 133    410

399 279 82    361

400 282 92    374

401 297 122    419

402 227 142    369

403 220 177    397

404 266 156    422

405 195 159    354

406 203 131    334

407 252 168    420

408 245 141     386

409 196 162     358

410 259 185     444

411 262 196     458

412 268 181     449

413 314 121     435

414 239 171     410

415 251 176     427

416 264 160     424

417 211 135     346

418 221 181     402

419 278 185     463

420 236 228     464

421 239 185     424

422 290 232     522

423 266 204     470

424 267 169     436

425 279 154     433

426 243 145     388

427 275 200    475

428 272 159    431

429 245 126    371

430 248 218    466

431 289 232    521

432 267 155    422

433 249 170    419

434 244 166    410

435 296 210    506

436 288 216    504

437 273 197    470

438 236 192    428

439 259 335    594

440 306 162    468

441 192 121    313

442 223 184    407

443 295 196    491

444 254 235    489

445 289 241    530

446 316 170    486

447 303 211    514

448 264 206    470

449 322 247    569

450 187 143    330

451 266 228    494

452 286 196    482

453 198 183    381

454 231 185    416

455 225 210    435

456 249 238    487

457 253 182    435

458 267 248    515

459 272 219    491

460 288 203    491

461 338 224    562

462 234 156    390

463 240 199    439

464 275 169    444

465 187 149    336

466 227 241    468

467 278 224    502

468 236 193    429

469 239 136    375

470 266 188    454

471 292 195    487

472 284 174    458

473 314 193    507

474 194 181    375

475 242 163    405

476 270 210    480

477 169 135    304

478 284 222    506

479 245 219    464

480 257 191    448

481 216 178    394

482 250 246    496

483 257 250    507

484 272 222    494

485 252 210    462

486 211 170    381

487 261 311    572

488 288 185    473

489 241 167    408

490 227 181    408

491 305 227    532

492 231 209    440

493 270 242    512

494 297 261    558

495 263 237    500

496 258 218    476

497 461 259    720

498 174 217    391

499 230 271    501

500 256 181    437

501 221 223    444

502 205 236    441

503 252 278    530

504 237 275    512

505 216 230    446

506 251 266    517

507 261 192    453

508 271 189    460

509 312 209    521

510 192 286    478

511 269 228    497

512 288 184    472

513 142 124    266

514 258 231    489

515 249 247    496

516 251 266    517

517 251 243    494

518 226 234    460

519 261 212    473

520 280 221    501

521 338 238    576

522 184 164    348

523 224 256    480

524 286 143    429

525 194 157    351

526 208 176    384

527 279 284    563

528 237 191    428

529 176 140    316

530 270 177    447

531 224 246    470

532 250 193    443

533 248 202    450

534 108 198    306

535 163 247    410

536 135 134    269

#Results #1: Summary statistics

library(fBasics);library(forestmangr)

Diethelm Wuertz, Tobias Setz and Yohan Chalabi (2020). fBasics: Rmetrics -

Markets and Basic Statistics. R package version 3042.89.1.

https://CRAN.R-project.org/package=fBasics

citation("forestmangr")

Sollano Rabelo Braga, Marcio Leles Romarco de

Oliveira and Eric Bastos Gorgens (2021).

forestmangr: Forest Mensuration and Management. R

package version 0.9.3.

https://CRAN.R-project.org/package=forestmangr

forestmangr::round_df(fBasics::basicStats(MD), digits=2)

|  | PMN | PMA | TotalMD |
|---|---|---|---|
| nobs | 536.00 | 536.00 | 536.00 |
| NAs | 0.00 | 0.00 | 0.00 |
| Minimum | 3.00 | 0.00 | 7.00 |
| Maximum | 813.00 | 335.00 | 869.00 |
| 1. Quartile | 247.00 | 32.00 | 327.00 |
| 3. Quartile | 347.25 | 135.25 | 437.25 |

Mean          296.11    83.59    379.70

Median        286.00    50.00    388.00

Sum          158714.00 44805.00 203519.00

SE Mean        3.45     3.25     4.11

LCL Mean       289.33    77.21    371.62

UCL Mean       302.89    89.98    387.78

Variance       6389.99  5665.03  9058.60

Stdev         79.94    75.27    95.18

Skewness       1.00     1.06    -0.02

Kurtosis       5.14    -0.11     2.34

#normality test

library(foreach); citation("foreach")

 Microsoft and Steve Weston (2020). foreach: Provides Foreach Looping Construct. R package version

1.5.1.

 https://CRAN.R-project.org/package=foreach

library(nortest); citation("nortest")

 Juergen Gross and Uwe Ligges (2015). nortest: Tests for Normality. R package version 1.0-4.

https://CRAN.R-project.org/package=nortest

```
ad<-foreach(i=1:3) %do% ad.test(MD[,i]) # Null: Normal distributed

cvm<-foreach(i=1:3) %do% cvm.test(MD[,i]) # Null: Normal distributed

KS<-foreach(i=1:3) %do% lillie.test(MD[,i]) # Null: Normal distribution
```

```
ad[1]; cvm[1]; KS[1]
```

Anderson-Darling normality test

A = 2.5753, p-value = 1.665e-06

Cramer-von Mises normality test

W = 0.46434, p-value = 5.952e-06

Lilliefors (Kolmogorov-Smirnov) normality test

D = 0.063404, p-value = 2.391e-05

```
ad[2]; cvm[2]; KS[2]
```

Anderson-Darling normality test

A = 7.6022, p-value < 2.2e-16

Cramer-von Mises normality test

W = 1.4083, p-value = 7.37e-10

Lilliefors (Kolmogorov-Smirnov) normality test

D = 0.094654, p-value = 2.012e-12

ad[3]; cvm[3]; KS[3]

Anderson-Darling normality test

A = 36.156, p-value < 2.2e-16

Cramer-von Mises normality test

W = 6.9231, p-value = 7.37e-10

Lilliefors (Kolmogorov-Smirnov) normality test

D = 0.23466, p-value < 2.2e-16

#seasonality

library(seastests)

citation("seastests")

Daniel Ollech (2019). seastests: Seasonality Tests. R

package version 0.14.2. https://CRAN.R-project.org/package=seastests

foreach(i=1:3) %do% seastests::isSeasonal(MD[,i], freq=12, "qs")

[[1]] [1] TRUE

[[2]] [1] TRUE

[[3]] [1] TRUE

foreach(i=1:3) %do% seastests::isSeasonal(MD[,i], freq=12, "fried")

[[1]] [1] TRUE

[[2]] [1] TRUE

[[3]] [1] TRUE

summary(seastests::wo(MD[,1], freq=12))

Test used:  WO

Test statistic:  1

P-value:  0 0 0

The WO - test identifies seasonality

summary(seastests::wo(MD[,2], freq=12))

Test used:  WO

Test statistic:  1

P-value:  6.081147e-11 1.012118e-07 3.314662e-06

The WO - test identifies seasonality

summary(seastests::wo(MD[,3], freq=12))

Test used:  WO

Test statistic:  1

P-value:  0 0 0

The WO - test identifies seasonality

#Nonlinearity

library(nonlinearTseries); citation("nonlinearTseries")

To cite package 'nonlinearTseries' in publications

use:

Constantino A. Garcia (2021). nonlinearTseries:

Nonlinear Time Series Analysis. R package version

0.2.11.

https://CRAN.R-project.org/package=nonlinearTseries

nonlinearity<- foreach(i=1:3) %do% nonlinearityTest(MD[,i],verbose=FALSE)

nonlinearity

[[1]]

[[1]]$Terasvirta #Null hypothesis: linearity in ``mean'

	Teraesvirta Neural Network Test

data:  ts(time.series)

X-squared = 4.9388, df = 2, p-value = 0.08464

[[1]]$White #Null hypothesis: linearity in ``mean'

	White Neural Network Test

data:  ts(time.series)

X-squared = 9.4799, df = 2, p-value = 0.008739

[[1]]$Keenan

[[1]]$Keenan$test.stat

[1] 0.1480264

[[1]]$Keenan$df1

[1] 1

[[1]]$Keenan$df2

[1] 508

[[1]]$Keenan$p.value

[1] 0.7005896

[[1]]$Keenan$order

[1] 13

[[1]]$McLeodLi

[[1]]$McLeodLi$p.values

 [1] 0 0 0 0 0 0 0 0 0 0 0 0 0 0 0 0 0 0 0 0 0 0 0 0 0 0 0 0 0 0

[[1]]$Tsay

[[1]]$Tsay$test.stat

[1] 2.705443

[[1]]$Tsay$p.value

[1] 8.701031e-12

[[1]]$Tsay$order

[1] 13

[[1]]$TarTest

[[1]]$TarTest$percentiles

[1] 24.8566 75.1434

[[1]]$TarTest$test.statistic

[1] 20.97269

[[1]]$TarTest$p.value

[1] 0.3417607

[[2]]

[[2]]$Terasvirta

Teraesvirta Neural Network Test

data:  ts(time.series)

X-squared = 130.32, df = 2, p-value < 2.2e-16

[[2]]$White

White Neural Network Test

data:  ts(time.series)

X-squared = 132.31, df = 2, p-value < 2.2e-16

[[2]]$Keenan

[[2]]$Keenan$test.stat

[1] 5.27641

[[2]]$Keenan$df1

[1] 1

[[2]]$Keenan$df2

[1] 494

[[2]]$Keenan$p.value

[1] 0.02203366

[[2]]$Keenan$order

[1] 20

[[2]]$McLeodLi

[[2]]$McLeodLi$p.values

 [1] 0 0 0 0 0 0 0 0 0 0 0 0 0 0 0 0 0 0 0 0 0 0 0 0 0 0 0 0

[[2]]$Tsay

[[2]]$Tsay$test.stat

[1] 7.975732

[[2]]$Tsay$p.value

[1] 1.718763e-55

[[2]]$Tsay$order

[1] 20

[[2]]$TarTest

[[2]]$TarTest$percentiles

[1] 25 75

[[2]]$TarTest$test.statistic

[1] 57.45707

[[2]]$TarTest$p.value

[1] 0.0008712318

[[3]]

[[3]]$Terasvirta

Teraesvirta Neural Network Test

data:  ts(time.series)

X-squared = 37.342, df = 2, p-value = 7.785e-09

[[3]]$White

    White Neural Network Test

data:  ts(time.series)

X-squared = 49.997, df = 2, p-value = 1.391e-11

[[3]]$Keenan

[[3]]$Keenan$test.stat

[1] 2.439916

[[3]]$Keenan$df1

[1] 1

[[3]]$Keenan$df2

[1] 508

[[3]]$Keenan$p.value

[1] 0.1189053

[[3]]$Keenan$order

[1] 13

[[3]]$McLeodLi

[[3]]$McLeodLi$p.values

 [1] 0 0 0 0 0 0 0 0 0 0 0 0 0 0 0 0 0 0 0 0 0 0 0 0 0 0 0 0

[[3]]$Tsay

[[3]]$Tsay$test.stat

[1] 2.050339

[[3]]$Tsay$p.value

[1] 1.035803e-06

```
[[3]]$Tsay$order

[1] 13

[[3]]$TarTest

[[3]]$TarTest$percentiles

[1] 24.8566 75.1434

[[3]]$TarTest$test.statistic

[1] 32.01621

[[3]]$TarTest$p.value

[1] 0.05321551
```

#Stationarity tests

library(aTSA); citation("aTSA")

      Debin Qiu (2015). aTSA: Alternative Time Series Analysis. R

      package version 3.1.2. https://CRAN.R-project.org/package=aTSA

adf.test(MD[,1])

Augmented Dickey-Fuller Test

alternative: stationary

Type 1: no drift no trend

   lag  ADF p.value

[1,]  0 -2.189  0.0287

[2,]  1 -1.508  0.1400

[3,]  2 -1.051  0.3033

[4,]  3 -0.929  0.3469

[5,]  4 -0.798  0.3936

[6,]  5 -0.685  0.4341

Type 2: with drift no trend

   lag  ADF p.value

[1,]  0 -9.46  0.0100

[2,]  1 -6.73  0.0100

[3,]  2 -4.67  0.0100

[4,]  3 -4.03  0.0100

[5,]  4 -3.50  0.0100

[6,]  5 -3.06  0.0313

Type 3: with drift and trend

   lag   ADF p.value

[1,]  0 -9.85  0.0100

[2,]  1 -7.09  0.0100

[3,]  2 -5.04  0.0100

[4,]  3 -4.40  0.0100

[5,]  4 -3.89  0.0140

[6,]  5 -3.48  0.0441

----

Note: in fact, p.value = 0.01 means p.value <= 0.01

adf.test(MD[,2])

Augmented Dickey-Fuller Test

alternative: stationary

Type 1: no drift no trend

   lag    ADF p.value

[1,]  0 -2.906   0.010

[2,]  1 -1.640  0.097

[3,]  2 -0.890  0.361

[4,]  3 -0.186  0.590

[5,]  4 0.111  0.676

[6,]  5 0.352  0.745

Type 2: with drift no trend

    lag   ADF p.value

[1,]  0 -4.605  0.0100

[2,]  1 -2.928  0.0444

[3,]  2 -2.025  0.3181

[4,]  3 -1.304  0.5941

[5,]  4 -1.051  0.6836

[6,]  5 -0.863  0.7503

Type 3: with drift and trend

    lag   ADF p.value

[1,]  0 -9.40  0.0100

[2,]  1 -6.23  0.0100

[3,]  2 -4.49  0.0100

[4,]  3 -3.14  0.0981

[5,]  4 -2.70  0.2826

[6,]  5 -2.41  0.4045

----

Note: in fact, p.value = 0.01 means p.value <= 0.01

adf.test(MD[,3])

Augmented Dickey-Fuller Test

alternative: stationary

Type 1: no drift no trend

    lag   ADF p.value

[1,]  0 -2.030  0.0428

[2,]  1 -1.288  0.2186

[3,]  2 -0.788  0.3973

[4,]  3 -0.570  0.4751

[5,]  4 -0.442  0.5166

[6,]  5 -0.297  0.5584

Type 2: with drift no trend

    lag   ADF p.value

[1,]   0 -9.73  0.0100

[2,]   1 -6.86  0.0100

[3,]   2 -4.88  0.0100

[4,]   3 -4.10  0.0100

[5,]   4 -3.71  0.0100

[6,]   5 -3.39  0.0123

Type 3: with drift and trend

   lag    ADF p.value

[1,]   0 -11.25  0.0100

[2,]   1  -7.90  0.0100

[3,]   2  -5.43  0.0100

[4,]   3  -4.46  0.0100

[5,]   4  -3.96  0.0106

[6,]   5  -3.51  0.0410

----

Note: in fact, p.value = 0.01 means p.value <= 0.01

kpss.test(MD[,1])

KPSS Unit Root Test

alternative: nonstationary

Type 1: no drift no trend

 lag stat p.value

  5 7.22   0.01

-----

 Type 2: with drift no trend

 lag stat p.value

  5 1.85   0.01

-----

 Type 1: with drift and trend

 lag  stat p.value

  5 0.954   0.01

-----------

Note: p.value = 0.01 means p.value <= 0.01

   : p.value = 0.10 means p.value >= 0.10

kpss.test(MD[,2])

KPSS Unit Root Test

alternative: nonstationary

Type 1: no drift no trend

 lag stat p.value

  5  3.3   0.01

-----

 Type 2: with drift no trend

 lag stat p.value

  5 3.89   0.01

-----

 Type 1: with drift and trend

 lag stat p.value

  5 1.38   0.01

-----------

Note: p.value = 0.01 means p.value <= 0.01

   : p.value = 0.10 means p.value >= 0.10

kpss.test(MD[,3])

KPSS Unit Root Test

alternative: nonstationary

Type 1: no drift no trend

 lag stat p.value

  5 6.29   0.01

-----

 Type 2: with drift no trend

 lag stat p.value

  5 2.16   0.01

-----

 Type 1: with drift and trend

 lag  stat p.value

  5 0.655   0.01

-----------

Note: p.value = 0.01 means p.value <= 0.01

   : p.value = 0.10 means p.value >= 0.10

pp.test(MD[,1])

Phillips-Perron Unit Root Test

alternative: stationary

Type 1: no drift no trend

 lag Z_rho p.value

  6 -2.62  0.355

-----

 Type 2: with drift no trend

 lag Z_rho p.value

  6  -145   0.01

-----

 Type 3: with drift and trend

 lag Z_rho p.value

  6  -156   0.01

--------------

Note: p-value = 0.01 means p.value <= 0.01

pp.test(MD[,2])

Phillips-Perron Unit Root Test

alternative: stationary

Type 1: no drift no trend

 lag Z_rho p.value

  6 -4.61   0.19

-----

 Type 2: with drift no trend

 lag Z_rho p.value

  6  -18  0.0203

-----

 Type 3: with drift and trend

 lag Z_rho p.value

  6  -152   0.01

--------------

Note: p-value = 0.01 means p.value <= 0.01

pp.test(MD[,3])

Phillips-Perron Unit Root Test

alternative: stationary

Type 1: no drift no trend

 lag Z_rho p.value

  6 -1.97  0.408

-----

 Type 2: with drift no trend

 lag Z_rho p.value

  6  -151   0.01

-----

 Type 3: with drift and trend

 lag Z_rho p.value

  6  -227   0.01

--------------

Note: p-value = 0.01 means p.value <= 0.01

#Estimate order of integration (d) - must be at most I(1), as diff(timeseries)=I(0)

library(LongMemoryTS); citation("LongMemoryTS")

        Christian Leschinski (2019). LongMemoryTS: Long

        Memory Time Series. R package version 0.1.0.

https://CRAN.R-project.org/package=LongMemoryTS

```
m<-floor(1+536^0.8)

m

[1] 153

gph(X=MD[,1], m=m)

[1] 0.3898275

gph(X=MD[,2], m=m)

[1] 0.6492232

gph(X=MD[,3], m=m)

[1] 0.4448909

#Long-Memory

acf(MD[,1], lag=100)

acf(MD[,2], lag=100)

acf(MD[,3], lag=100)

# Calculating Hurst

library(pracma); citation("pracma")
```

Hans W. Borchers (2021). pracma: Practical Numerical Math Functions. R package version 2.3.3.

https://CRAN.R-project.org/package=pracma

# hurstexp()

library(tsfeatures); citation("tsfeatures")

Rob Hyndman, Yanfei Kang, Pablo Montero-Manso, Thiyanga Talagala, Earo Wang, Yangzhuoran

Yang and

Mitchell O'Hara-Wild (2020). tsfeatures: Time Series Feature Extraction. R package version 1.0.2.

https://CRAN.R-project.org/package=tsfeatures

#hurst

hurst(MD[,1])

hurst

0.9294749

pracma::hurstexp(MD[,1])

Simple R/S Hurst estimation:          0.8289719

Corrected R over S Hurst exponent:   1.030994

Empirical Hurst exponent:          1.13147

Corrected empirical Hurst exponent:  1.084217

Theoretical Hurst exponent:          0.5455964

hurst(MD[,2])

hurst

0.9957688

pracma::hurstexp(MD[,2])

Simple R/S Hurst estimation:        0.8589667

Corrected R over S Hurst exponent:   1.098318

Empirical Hurst exponent:        1.204543

Corrected empirical Hurst exponent: 1.166805

Theoretical Hurst exponent:        0.5455964

hurst(MD[,3])

hurst

0.9214155

pracma::hurstexp(MD[,3])

Simple R/S Hurst estimation:        0.7708484

Corrected R over S Hurst exponent:   1.004789

Empirical Hurst exponent:        1.262727

Corrected empirical Hurst exponent: 1.217228

Theoretical Hurst exponent:        0.5455964

#Structural Breaks

library(strucchange); citation("strucchange")

# Significant testing: Empirical fluctuation processes: Null hypothesis: No structural change

```
sctest( efp(MD[,1]~1, type="OLS-CUSUM"))
```

    OLS-based CUSUM test

data:  efp(MD[, 1] ~ 1, type = "OLS-CUSUM")

S0 = 4.5812, p-value < 2.2e-16

```
sctest( efp(MD[,1]~1, type="OLS-MOSUM"))
```

    OLS-based MOSUM test

data:  efp(MD[, 1] ~ 1, type = "OLS-MOSUM")

M0 = 4.6639, p-value = 0.01

```
sctest( efp(MD[,1]~1, type="Rec-CUSUM"))
```

    Recursive CUSUM test

data:  efp(MD[, 1] ~ 1, type = "Rec-CUSUM")

S = 5.3071, p-value < 2.2e-16

sctest( efp(MD[,1]~1, type="Rec-MOSUM"))

    Recursive MOSUM test

data:  efp(MD[, 1] ~ 1, type = "Rec-MOSUM")

M = 4.754, p-value = 0.01

sctest( efp(MD[,2]~1, type="OLS-CUSUM"))

    OLS-based CUSUM test

data:  efp(MD[, 2] ~ 1, type = "OLS-CUSUM")

S0 = 9.6287, p-value < 2.2e-16

sctest( efp(MD[,2]~1, type="OLS-MOSUM"))

    OLS-based MOSUM test

data:  efp(MD[, 2] ~ 1, type = "OLS-MOSUM")

M0 = 5.9073, p-value = 0.01

sctest( efp(MD[,2]~1, type="Rec-CUSUM"))

        Recursive CUSUM test

data:  efp(MD[, 2] ~ 1, type = "Rec-CUSUM")

S = 6.661, p-value < 2.2e-16

sctest( efp(MD[,2]~1, type="Rec-MOSUM"))

        Recursive MOSUM test

data:  efp(MD[, 2] ~ 1, type = "Rec-MOSUM")

M = 8.702, p-value = 0.01

sctest( efp(MD[,3]~1, type="OLS-CUSUM"))

OLS-based CUSUM test

data:  efp(MD[, 3] ~ 1, type = "OLS-CUSUM")

S0 = 5.4535, p-value < 2.2e-16

sctest( efp(MD[,3]~1, type="OLS-MOSUM"))

OLS-based MOSUM test

data:  efp(MD[, 3] ~ 1, type = "OLS-MOSUM")

M0 = 5.376, p-value = 0.01

sctest( efp(MD[,3]~1, type="Rec-CUSUM"))

Recursive CUSUM test

data:  efp(MD[, 3] ~ 1, type = "Rec-CUSUM")

S = 6.6661, p-value < 2.2e-16

```
sctest( efp(MD[,3]~1, type="Rec-MOSUM"))
```

        Recursive MOSUM test

data:  efp(MD[, 3] ~ 1, type = "Rec-MOSUM")

M = 6.1793, p-value = 0.01

# Periodicities

library(tseries);library(forecast)

citation("tseries")

        Adrian Trapletti and Kurt Hornik (2020). tseries: Time Series

        Analysis and Computational Finance. R package version 0.10-48.

PMN<-ts(MD[,1], start=c(1976,5), frequency=12)

PMA<-ts(MD[,2], start=c(1976,5), frequency=12)

TotalMD<-ts(MD[,3], start=c(1976,5), frequency=12)

# STL: Seasonal

```
PMNstl<-stl(PMN, s.window=12)
```

```
PMAstl<- stl(PMA, s.window=12)
```

```
TotalMDstl<-stl(TotalMD, s.window=12)
```

```
plot(PMNstl$time.series[,1], ylab=""); Hmisc::minor.tick(10); #legend("topleft", legend=c("PMN
Applications"))
```

```
plot(PMAstl$time.series[,1], ylab=""); Hmisc::minor.tick(10); #legend("topleft", legend=c("PMA
Applications"))
```

```
plot(TotalMDstl$time.series[,1], ylab=""); Hmisc::minor.tick(10); #legend("topleft", legend=c("TotalMD
Applications"))
```

```
library(forecast); citation("forecast")
```

Hyndman R, Athanasopoulos G, Bergmeir C, Caceres G, Chhay L, O'Hara-Wild M, Petropoulos F, Razbash S, Wang E, Yasmeen

F (2021). _forecast: Forecasting functions for time series and linear models_. R package version 8.14, <URL:https://pkg.robjhyndman.com/forecast/>.

Hyndman RJ, Khandakar Y (2008). "Automatic time series forecasting: the forecast package for R." _Journal of Statistical Software_, *26*(3), 1-22.

<URL:https://www.jstatsoft.org/article/view/v027i03>.

```
findfrequency(PMN)/12
```

> [1] 1

```
findfrequency(PMA)/12
```

> [1] 0.3333333

```
findfrequency(TotalMD)/12
```

> [1] 0.25

```
library(WaveletComp); citation("WaveletComp")
```

> Angi Roesch and Harald Schmidbauer (2018). WaveletComp: Computational Wavelet Analysis.

> R package version 1.1. https://CRAN.R-project.org/package=WaveletComp

```
monthyear <- seq(as.Date("1976-05-01"), as.Date("2020-12-31"), by = "month")

monthyear <- strftime(monthyear, format = "%b %Y")

#Spectral domain

PMNwl<- analyze.wavelet(data.frame(MD),"PMN", dt=1/12, dj=0.01)

PMAwl<- analyze.wavelet(data.frame(MD),"PMA", dt=1/12, dj=0.01)

TotalMDwl<- analyze.wavelet(data.frame(MD),"TotalMD", dt=1/12, dj=0.01)
```

```
wt.image(PMNwl,  periodlab = "Period (Years)", timelab = "Month /Year", spec.time.axis = list(at =
1:length(monthyear), labels = monthyear))

wt.image(PMAwl,  periodlab = "Period (Years)", timelab = "Month /Year", spec.time.axis = list(at =
1:length(monthyear), labels = monthyear))

wt.image(TotalMDwl,  periodlab = "Period (Years)", timelab = "Month /Year", spec.time.axis = list(at =
1:length(monthyear), labels = monthyear))

wt.avg(PMNwl); wt.avg(PMAwl); wt.avg(TotalMDwl)
```

```
#RMAF: Long trends

library(rmaf); citation("rmaf")
```

To cite package 'rmaf' in publications use:

 Debin Qiu (2015). rmaf: Refined Moving Average Filter. R package

 version 3.0.1. https://CRAN.R-project.org/package=rmaf

```
PMNrmaf<-ma.filter(MD$PMN, seasonal = TRUE, period = 12, plot=TRUE)

PMArmaf<-ma.filter(MD$PMA, seasonal = TRUE, period = 12, plot=TRUE)
```

```
TotalMDrmaf<-ma.filter(MD$TotalMD, seasonal = TRUE, period = 12, plot=TRUE)
```

```
library(pracma); citation("pracma")
```

Hans W. Borchers (2021). pracma: Practical Numerical Math Functions. R

package version 2.3.3. https://CRAN.R-project.org/package=pracma

```
library(zoo);citation("zoo")
```

To cite zoo in publications use:

Achim Zeileis and Gabor Grothendieck (2005). zoo: S3

Infrastructure for Regular and Irregular Time Series.

Journal of Statistical Software, 14(6), 1-27.

doi:10.18637/jss.v014.i06

```
findpeaks(TotalMDrmaf[,2], npeaks=200, threshold=0, minpeakheight=440)
```

       [,1] [,2] [,3] [,4]

 [1,] 437.2046  164  147  165

 [2,] 438.8526  168  165  169

 [3,] 439.0179  170  169  171

 [4,] 439.4210  174  171  175

 [5,] 439.1110  178  175  179

 [6,] 439.4591  183  179  184

 [7,] 440.5705  185  184  187

 [8,] 443.6578  192  187  196 #Peak 1

 [9,] 432.6743  197  196  202

[10,] 434.9404  426  417  427

[11,] 436.3103  429  427  431

[12,] 444.5656  442  431  443

[13,] 453.3363  453  443  454

[14,] 461.1863  459  454  462

[15,] 464.8786  468  462  470

[16,] 467.0380  473  470  475

[17,] 465.4791  477  475  478

[18,] 467.8616  483  478  484 #Peak 2

[19,] 465.5575  485  484  488

[20,] 464.3752  489  488  491

[21,] 462.0179  494  491  500

[22,] 454.6743  501  500  503

[23,] 454.9585  504  503  506

[24,] 454.0728  508  506  535

TotalMDrmafts<-ts(TotalMDrmaf, frequency=12, start=c(1976,5))

as.yearmon(time(TotalMDrmafts)[192]) #Peak #1:

[1] "Apr 1992"

as.yearmon(time(TotalMDrmafts)[483]) #Peak #2:

[1] "July 2016"

as.yearmon(time(TotalMDrmafts)[483]) - as.yearmon(time(TotalMDrmafts)[192])

plot(TotalMDrmafts, type="l");abline(v=as.yearmon(time(TotalMDrmafts)[192]),

col="red");abline(v=as.yearmon(time(TotalMDrmafts)[483]), col="red");

#Final Plot

#plot(MA510rPMAr[,2], type="l", col="red", ylab="");abline(h=t[192],

v=as.yearmon(time(MA510rPMAr[,2])[192]), col="red"); abline(h=MA510rPMAr[,2][483],

v=as.yearmon(time(MA510rPMAr[,2])[483]), col="red")

```
#lines(MA510cPMAa[,2], type="l", col="blue");abline(h=t[197],

v=as.yearmon(time(MA510cPMAa[,2])[197]), col="blue"); abline(h=MA510cPMAa[,2]515],

v=as.yearmon(time(MA510cPMAa[,2])[515]), col="blue")

#abline(h=MA510rPMAr[,2][310], v=as.yearmon(time(MA510rPMAr[,2])[310]), col="red");

#abline(h=MA510cPMAa[,2][324], v=as.yearmon(time(MA510cPMAa[,2])[324]), col="blue")

#minor.tick(nx=10, ny=10)

#Peak-to-Peak Distance:

# 2nd peak to 1st peak: Oct 2012 [442.0573] - Apr 1992 [443.6578]

#as.yearmon(time(MA510rPMAr[,2])[438]) - as.yearmon(time(MA510rPMAr[,2])[192])

#20.5 #20 years and 5 months: Mar 2019 [470.3957] - Sept 1992 [432.8777]

#as.yearmon(time(MA510cPMAa[,2])[515]) - as.yearmon(time(MA510cPMAa[,2])[197])

#26.5 #26 years and 5 months

#Peak-to-Trough Distance:

#MA510rPMAr[,2][310]

#[1] 338.5266

#> as.yearmon(time(MA510rPMAr[,2])[310]);

#[1] "Feb 2002"
```

#as.yearmon(time(MA510rPMAr[,2])[310])-as.yearmon(time(MA510rPMAr[,2])[192])

#[1] 9.833333

# MA510cPMAa[,2][324]

#[1] 340.5898

#as.yearmon(time(MA510cPMAa[,2])[324])

#[1] "Apr 2003"

#structural breaks

library(strucchange); citation("strucchange")

 Achim Zeileis, Friedrich Leisch, Kurt Hornik and Christian Kleiber (2002).

 strucchange: An R Package for Testing for Structural Change in Linear

 Regression Models. Journal of Statistical Software, 7(2), 1-38. URL

 http://www.jstatsoft.org/v07/i02/

 Achim Zeileis, Christian Kleiber, Walter Kraemer and Kurt Hornik (2003).

 Testing and Dating of Structural Changes in Practice. Computational

 Statistics & Data Analysis, 44, 109-123.

```
PMN<-ts(MD[,1], start=c(1976,5), frequency=12)

breakdates(confint(breakpoints(PMN ~ 1)))

plot(PMN); lines (confint(breakpoints(PMN ~ 1)))
```

```
        2.5 % breakpoints   97.5 %

1 1982.917    1983.083 1983.250

2 1996.667    1996.917 1997.583

3 2002.917    2003.667 2004.333
```

```
PMA<-ts(MD[,2], start=c(1976,5), frequency=12)

breakdates(confint(breakpoints(PMA ~ 1)))

plot(PMA); lines (confint(breakpoints(PMA ~ 1)))
```

```
        2.5 % breakpoints   97.5 %

1 1982.917    1983.083 1983.250

2 2005.083    2005.333 2005.417

3 2011.750    2012.000 2012.250
```

```
TotalMD<-ts(MD[,3], start=c(1976,5), frequency=12)

breakdates(confint(breakpoints(TotalMD ~ 1)))

plot(TotalMD); lines (confint(breakpoints(TotalMD ~ 1)))

 2.5 % breakpoints  97.5 %

1 1982.917    1983.083 1983.250

2 1996.583    1996.917 1997.917

3 2009.833    2010.333 2010.500
```

# EEMD / CEEMD

```
library(Rlibeemd); citation("Rlibeemd")
```

Helske J, Luukko P (2018). _Rlibeemd: Ensemble empirical mode

decomposition (EEMD) and its complete variant (CEEMDAN)_. R

package version 1.4.1, <URL:

https://github.com/helske/Rlibeemd>.

Luukko PJ, Helske J, Räsänen E (2016). "Introducing libeemd: A

program package for performing the ensemble empirical mode

```
#CEEMDAN

PMNceemdan <-ceemdan(MD$PMN); PMAceemdan <-ceemdan(MD$PMA); TotalMDceemdan <-

ceemdan(MD$TotalMD)

X11(); plot( ts(PMNceemdan[, 6], frequency=12, start=c(1976,5)), ylab="PMN"); Hmisc::minor.tick(10)

X11(); plot( ts(PMAceemdan[, 6], frequency=12, start=c(1976,5)), ylab="PMA" ); Hmisc::minor.tick(10)

X11(); plot( ts(TotalMDceemdan[, 6], frequency=12, start=c(1976,5)), ylab="TotalMD" );

Hmisc::minor.tick(10)

#NOT USED

#EEMD

#PMNeemd <-eemd(PMN); PMAeemd <-eemd(PMA); TotalMDeemd <-eemd(TotalMD)

#X11();plot(PMNeemd) #total

#X11();plot(PMAeemd) #total

#X11();plot(TotalMDeemd) #total
```

```
#PMNeemd_low_frequencies<- ts (rowSums(PMNeemd[, 4:ncol(PMNeemd)]), frequency=12,
start=c(1976,5))

#PMAeemd_low_frequencies<- ts (rowSums(PMAeemd[, 4:ncol(PMAeemd)]), frequency=12,
start=c(1976,5))

#TotalMD_low_frequencies<- ts (rowSums(TotalMDeemd[, 4:ncol(TotalMDeemd)]), frequency=12,
start=c(1976,5))

#NOT USED

#MODWT

#library(wavelets);citation("wavelets")

#       Eric Aldrich (2020). wavelets: Functions

#       for Computing Wavelet Filters, Wavelet

#       Transforms and Multiresolution Analyses. R

#       package version 0.3-0.2.

#       https://CRAN.R-project.org/package=wavelets

#PMNmodwt<-wavelets::modwt( as.numeric(PMN) , filter="haar", n.levels=9); plot.modwt( PMNmodwt
)

#PMAmodwt<-wavelets::modwt( as.numeric(PMA) , filter="haar", n.levels=9); plot.modwt( PMAmodwt
)
```

```
#TotalMDmodwt<-wavelets::modwt( as.numeric(TotalMD) , filter="haar", n.levels=9); plot.modwt(
TotalMDmodwt )
```

#END